\documentstyle[aps,epsfig]{revtex}
\newcommand{\lsim}{\mathrel{\rlap{\lower4pt\hbox{\hskip0pt$\sim$}}
\raise1pt\hbox{$<$}}}
\newcommand{\gsim}{\mathrel{\rlap{\lower4pt\hbox{\hskip0pt$\sim$}}
\raise1pt\hbox{$>$}}}
\newcommand{\sfrac}[2]{\mbox{\footnotesize $\frac{#1}{#2}$}}
\begin{document}
\twocolumn

\title{Selected nucleon form factors and a composite scalar diquark}
%
\author{J.C.R. Bloch, C.D. Roberts and S.M. Schmidt \vspace*{0.2em}}
\address{Physics Division, Argonne National Laboratory, Argonne IL
60439-4843\\[0.6\baselineskip] 
\parbox{140mm}{\rm \hspace*{1.0em} 
A covariant, composite scalar diquark, Fadde'ev amplitude model for the
nucleon is used to calculate pseudoscalar, isoscalar- and isovector-vector,
axial-vector and scalar nucleon form factors.  The last yields the nucleon
$\sigma$-term and on-shell $\sigma$-nucleon coupling.  The calculated form
factors are soft, and the couplings are generally in good agreement with
experiment and other determinations.  Elements in the
dressed-quark-axial-vector vertex that are not constrained by the
Ward-Takahashi identity contribute $\sim 20$\% to the magnitude of $g_A$.
The calculation of the nucleon $\sigma$-term elucidates the only unambiguous
means of extrapolating meson-nucleon couplings off the meson
mass-shell. \\[0.4\baselineskip]
Pacs Numbers: 24.85.+p, 14.20.Dh, 13.75.Gx, 13.75.-n}}
%
%
\maketitle



\section{Introduction}
Current generation experiments probe hadrons and their interactions on a
truly dynamical domain where symmetries alone are insufficient to
characterise them.  In this domain phenomenologically accurate
nucleon-nucleon potentials\cite{machleidt,vincent} and meson exchange
models\cite{harry} are keys in the interpretation of data.  These models are
tools via which the correlated quark exchange underlying hadron-hadron
interactions is realised as a sum of exchanges of elementary, meson-like
degrees of freedom,\footnote{The extent to which these degrees of freedom are
identified with the mesons of the strong interaction spectrum varies.  In one
boson exchange models\protect\cite{machleidt} the identification is close,
while in the Argonne series of potentials\cite{vincent} the short-range part
is interpreted as a purely phenomenological parametrisation.}
and their definition relies on meson-nucleon form factors that sensibly
provide short-range cutoffs in the integrals that arise in calculations.

These form factors are interpreted as a manifestation of the hadrons'
internal structure.  If this interpretation is realistic then they should be
calculable in models that reliably describe hadron structure.  This cannot
mean that models of hadron structure should exactly reproduce the
momentum-dependence and parameter values used in potential models.  In order
to be phenomenologically successful, all models have hidden degrees of
freedom, which make complicated a direct comparison between approaches.
However, one can expect semi-quantitative agreement, with large discrepancies
being harbingers of model artefacts and defects.

An additional complication is that the mesons of the strong interaction
spectrum are bound states and hence are only unambiguously defined on-shell;
i.e., at their pole position in a n-point vertex function.  Any reference to
an off-shell meson is {\it necessarily} model dependent.  Therefore the only
comparisons that can be model-independent are those between calculated
meson-baryon coupling constants and on-shell couplings inferred from
potential models because these comparisons do not involve the {\it ad hoc}
definition of an off-shell bound state.

Primarily for this reason, the comparison between calculated and
phenomenological form factors can only be qualitative and should employ more
than one off-shell extrapolation to provide reliable information.  In spite
of the ambiguities, however, the calculation of meson-baryon form factors is
an essential element of contemporary phenomenology.  For example, it can
expose difficulties in phenomenological interpretations, as is well
exemplified by the discussion of $\rho^0$-$\omega$ mixing and its
contribution to charge symmetry breaking in $NN$
potentials\cite{rhoomega},\footnote{It is an important and model-independent
result that vector-channel resonant quark exchange is described by a vacuum
polarisation: $\Pi_{\mu\nu}(k)$, that vanishes at
$k^2=0$\protect\cite{conrad,peteradelaide}.}
and also provide guidance in constraining meson exchange currents in light
nuclear systems\cite{peteradelaide,deuterium}.

The dominant meson-like degrees of freedom employed in potential models are
identified with the $\pi$, $\rho$, $\omega$ and a light scalar, $\sigma$.
Herein we calculate the associated meson-nucleon coupling constants and form
factors using a covariant nucleon model\cite{jacques}.  It is motivated by
quark-diquark solutions of a relativistic Fadde'ev
equation\cite{reg,ishii,raA} and while only retaining a scalar diquark
correlation is a limitation, the model's treatment of that as a nonpointlike,
confined composite is a significant beneficial feature.  That is illustrated
in its application to the calculation of nucleon electromagnetic form
factors\cite{jacques}, which semi-quantitatively describes the ratio $\mu_p
G_E^p(q^2)/G_M^p(q^2)$ recently observed at TJNAF\cite{HallA}.  In Sec.~II we
review the model.  Our results are described in the next four sections:
Sec.~\ref{secpiN}, $\pi NN$; Sec.~\ref{secVN}, $\omega NN$- and $\rho
NN$-like interactions; Sec.~\ref{secgA} explores the nucleon's axial-vector
current; and Sec.~\ref{secsig} focuses on the scalar-nucleon interaction.
Sec.~\ref{secSC} is a brief recapitulation and an appendix contains selected
formulae.

\section{Nucleon Model}
\label{Sect:Model}
We represent the nucleon as a three-quark bound state involving a
nonpointlike diquark correlation and write its Fadde'ev amplitude as
\begin{eqnarray}
\label{Psi}
\lefteqn{\Psi^{\tau}_\alpha(p_1,\alpha_1,\tau^1;p_2,\alpha_2,\tau^2;
                p_3,\alpha_3,\tau^3)   = } \\
&& \nonumber 
\varepsilon_{c_1 c_2 c_3}\,
\delta^{\tau \tau^3}\,\delta_{\alpha \alpha_3}\,\psi(p_1+p_2,p_3)\,
\Delta(p_1+p_2)\,
\Gamma_{\alpha_1 \alpha_2}^{\tau^1 \tau^2}(p_1,p_2) \,,
\end{eqnarray}
where $\varepsilon_{c_1 c_2 c_3}$ effects a singlet coupling of the quarks'
colour indices, $(p_i,\alpha_i,\tau^i)$ denote the momentum and the Dirac and
isospin indices for the $i$-th quark constituent, $\alpha$ and $\tau$ are
these indices for the nucleon itself, $\psi(\ell_1,\ell_2)$ is a
Bethe-Salpeter-like amplitude characterising the relative-momentum dependence
of the correlation between diquark and quark, $\Delta(K)$ describes the
propagation characteristics of the diquark, and
\begin{eqnarray}
\label{gdq}
\Gamma_{\alpha_1 \alpha_2}^{\tau^1 \tau^2}(p_1,p_2) & = &
(C i\gamma_5)_{\alpha_1 \alpha_2}\, (i\tau_2)^{\tau^1\tau^2}\,
\Gamma (p_1,p_2)
\end{eqnarray}
represents the momentum-dependence, and spin and isospin character of the
diquark correlation; i.e., it corresponds to a Bethe-Salpeter-like amplitude
for the diquark.  While complete antisymmetrisation is not explicit in
$\Psi$, it is exhibited in our calculations via the exchange of roles between
the dormant and diquark-participant quarks, and gives rise to diquark
``breakup'' contributions to the form factors.  This in not an afterthought,
it merely reflects the simple manner in which we choose to order and
elucidate our calculations.

With the form of $\Psi$ in Eq.~(\ref{Psi}), we retain in the quark-quark
scattering matrix only the contribution of the scalar diquark, which has the
largest correlation length~\cite{bsesep}: $\lambda_{0^+}:=1/m_{0^+} =
0.27\,$fm.  We saw as anticipated in Ref.~\cite{jacques} that the primary
defect of Eq.~(\ref{Psi}) is the omission of the axial-vector correlation
($\lambda_{1^+} \approx 0.8\,\lambda_{0^+}$).  Nevertheless the {\it Ansatz}
yielded much about the electromagnetic nucleon form factors that was
quantitatively reliable and qualitatively informative.  Hence, we employ it
again herein as an exploratory, intuition building tool.

The amplitude in Eq.~(\ref{Psi}) is fully determined with the specification
of the scalar functions:
\begin{eqnarray}
\label{littlepsi}
\psi(\ell_1,\ell_2) & = & \frac{1}{{\cal N}_\Psi}\,{\cal
F}(\ell^2/\omega_\psi^2)\,,\;\ell := \case{1}{3}\,\ell_1 -
\case{2}{3}\,\ell_2\,,\\ 
\Gamma(q_1,q_2) & = & 
 \frac{1}{{\cal N}_\Gamma}\,
        {\cal F}(q^2/\omega_\Gamma^2)\,,\;q:=
\case{1}{2}\,q_1-\case{1}{2}\,q_2\,, \\
\label{dprop}
\Delta(K)  & = & \frac{1}{m_\Delta^2}\,{\cal F}(K^2/\omega_\Gamma^2)\,,\\
{\cal F}(y) & = & \frac{1- {\rm e}^{-y}}{y}\,,
\end{eqnarray}
which introduces three parameters whose values were determined\cite{jacques}
in a least-squares fit to $G^p_E(q^2)$
\begin{equation}
\label{params}
\begin{array}{ccc}
\;\omega_\psi & \omega_\Gamma & m_\Delta \\\hline
 \;0.20  & 1.4   & 0.63
\end{array}
\end{equation}
all in GeV ($1/m_\Delta = 0.31\,$fm).\footnote{This modified value of
$\omega_\Gamma$ arises from correcting a minor computational error in the
calculations of Ref.~\protect\cite{jacques}.  In our Euclidean formulation:
$p\cdot q=\sum_{i=1}^4 p_i q_i$,
$\{\gamma_\mu,\gamma_\nu\}=2\,\delta_{\mu\nu}$, $\gamma_\mu^\dagger =
\gamma_\mu$, $\sigma_{\mu\nu}= \case{i}{2}[\gamma_\mu,\gamma_\nu]$, and
tr$_D[\gamma_5\gamma_\mu\gamma_\nu\gamma_\rho\gamma_\sigma]=
-4\,\epsilon_{\mu\nu\rho\sigma}$, $\epsilon_{1234}= 1$.}  ${\cal N}_\Psi$ and
${\cal N}_\Gamma$ are the {\it calculated} nucleon and $(ud)$ diquark
normalisation constants, which via the canonical definition ensure composite
electric charges of $1$ for the proton and $1/3$ for the diquark.  Current
conservation is manifest in this model.

An essential, additional element in the calculation of the electromagnetic
nucleon form factors is the dressed-quark propagator:
\begin{eqnarray}
\label{qprop}
S(p) & = & -i\gamma\cdot p\, \sigma_V(p^2) + \sigma_S(p^2)\\
\label{Sinv}
& = & \left[i \gamma\cdot p \, A(p^2) + B(p^2)\right]^{-1}\,.
\end{eqnarray}
While $S(p)$ can be obtained as a solution of the quark Dyson-Schwinger
equation (DSE)\cite{cdragw}, a phenomenologically efficacious algebraic
parametrisation has been determined in extensive studies of meson
properties\cite{mark,echaya} and we employ it herein:
\begin{eqnarray}
\label{ssm}
\bar\sigma_S(x) & =&  2\,\bar m \,{\cal F}(2 (x+\bar m^2))\\
&& \nonumber
+ {\cal F}(b_1 x) \,{\cal F}(b_3 x) \,
\left[b_0 + b_2 {\cal F}(\epsilon x)\right]\,,\\
\label{svm}
\bar\sigma_V(x) & = & \frac{1}{x+\bar m^2}\,
\left[ 1 - {\cal F}(2 (x+\bar m^2))\right]\,,
\end{eqnarray}
$x=p^2/\lambda^2$, $\bar m$ = $m/\lambda$, $\bar\sigma_S(x) =
\lambda\,\sigma_S(p^2)$ and $\bar\sigma_V(x) = \lambda^2\,\sigma_V(p^2)$.
The mass-scale, $\lambda=0.566\,$GeV, and parameter values
\begin{equation}
\label{tableA} 
\begin{array}{ccccc}
   \bar m& b_0 & b_1 & b_2 & b_3 \\\hline
   0.00897 & 0.131 & 2.90 & 0.603 & 0.185 
\end{array}\;,
\end{equation}
were fixed in a least-squares fit to light-meson observables\cite{mark}.
($\epsilon=10^{-4}$ in (\ref{ssm}) acts only to decouple the large- and
intermediate-$p^2$ domains.)  This algebraic parametrisation combines the
effects of confinement and dynamical chiral symmetry breaking with
free-particle behaviour at large spacelike $p^2$~\cite{echaya}.

\section{Pion Nucleon Coupling}
\label{secpiN}
The pion-nucleon current is
\begin{eqnarray}
\label{PiNcur}
J_\pi^j(P^\prime,P) & = & \bar u(P^\prime)\,\Lambda_\pi^j(q,P)\,u(P)\\
&=:& g_{\pi N N}(q^2)\,\bar u(P^\prime)\,
 i\tau^j\gamma_5  \,u(P)\,,
\end{eqnarray}
where the spinors satisfy: 
\begin{equation}
\label{spinors}
\gamma\cdot P \, u(P) = i M u(P)\,,\;
\bar u(P)\,\gamma\cdot P = i M \bar u(P)
\end{equation}
with the nucleon mass $M=0.94\,$GeV and $q=(P^\prime-P)$.

For an on-shell pion a calculation of the impulse approximation to $J_\pi^j$
requires only one additional element: $\Gamma^j_{\pi}(k;Q)$, the pion
Bethe-Salpeter amplitude, with $k$ the relative quark-antiquark momentum and
$Q$ the total momentum of the bound state.  It has the general form
\begin{eqnarray}
\label{genpibsa}
\Gamma_\pi^j(k;Q) & = &  \tau^j \gamma_5 \left[ i E_\pi(k;Q) + 
\gamma\cdot Q \,F_\pi(k;Q) \rule{0mm}{5mm}\right. \\
\nonumber
&+ & \left. \rule{0mm}{5mm} \gamma\cdot k \,k \cdot Q\, G_\pi(k;Q) 
+ \sigma_{\mu\nu}\,k_\mu Q_\nu \,H_\pi(k;Q) 
\right]
\end{eqnarray}
and is obtained as a solution of an homogeneous Bethe-Salpeter equation.

Using any truncation of the quark-antiquark scattering matrix that ensures
the preservation of the axial-vector Ward-Takahashi identity then, in the
chiral limit\cite{mrt98},
\begin{eqnarray}
\label{piamp}
E_\pi(k;Q=0) & = & \frac{1}{f_\pi} B_0(k^2)\,,
\end{eqnarray}
and $F_\pi$, $G_\pi$, $H_\pi$ satisfy similar relations involving $A_0(k^2)$.
Here $f_\pi$ is the pion decay constant and $A_0(k^2)$, $B_0(k^2)$ are the
dressed-quark propagator functions in Eq.~(\ref{Sinv}) calculated in the
chiral limit.  Since\cite{cdragw,echaya}
\begin{equation}
A(p^2) \neq  1\,,
\end{equation}
the identities involving $F_\pi$, $G_\pi$, $H_\pi$ entail that the pion
necessarily has pseudovector components, even in the chiral limit.  These
components are crucial at large pion energy; e.g., they are responsible for
the asymptotic $1/q^2$-behaviour of the electromagnetic pion form
factor\cite{mrpion}, however, for pion energy $\lsim 1\,$GeV they are
quantitatively unimportant, and Eq.~(\ref{genpibsa}) with Eq.~(\ref{piamp})
and $F_\pi=0=G_\pi=H_\pi$ provides a reliable approximation.

This fact is useful in phenomenological applications, and away from the
chiral limit an algebraic parametrisation has been developed\cite{mark,hawes}
to be used in concert with Eqs.~(\ref{ssm},\ref{svm}):
\begin{equation}
\label{actualpi}
E_\pi(k;Q) = \frac{1}{f_\pi} \,B_\pi(k^2)\,,
\end{equation}
where $B_\pi(k^2)$ is obtained from Eqs.~(\ref{Sinv}-\ref{svm})
with\cite{misha}
\begin{equation}
\bar m \to 0\,,\;
b_0 \to b_0^\pi = 0.204\,.
\end{equation}
This form of dressed-quark propagator and pion Bethe-Salpeter amplitude
yields (quoted with GeV as the base unit)
\begin{equation}
\label{oldresults}
\begin{array}{l|cccc}
  & f_\pi & m_\pi 
& \langle \bar q q\rangle^{1\,{\rm GeV}^2}_0
& \langle \bar q q\rangle^{1\,{\rm GeV}^2}_\pi \\\hline
{\rm Calc.} & 0.0924 & 0.141 & (0.221)^3 & (0.257)^3\\
{\rm Obs.\protect\cite{pdg,mr97}}  & 0.0924 & 0.138 & 
        (0.241)^3 & (0.245)^3
\end{array}
\end{equation}

The (on-shell) Bethe-Salpeter amplitude is sufficient to calculate the
pion-nucleon coupling.  However, to calculate the form factor we must specify
an off-shell extrapolation of $E_\pi(k;Q)$; i.e., a functional dependence for
$Q^2\neq -m_\pi^2$.  Two obvious {\it Ans\"atze} are
\begin{figure}
\centering{\ \epsfig{figure=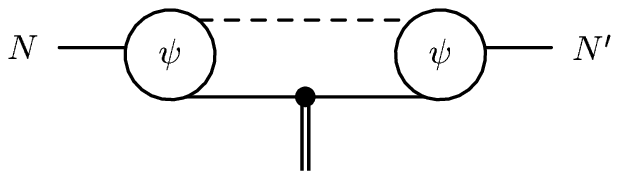,height=1.5cm}}

\vspace*{\baselineskip}

\centering{\ \epsfig{figure=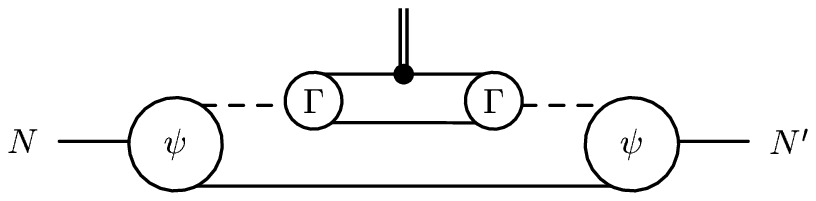,height=1.5cm}}

\vspace*{\baselineskip}

\centering{\ \epsfig{figure=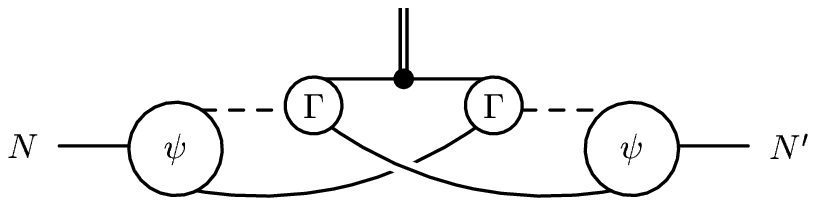,height=1.5cm}}

\vspace*{\baselineskip}

\centering{\ \epsfig{figure=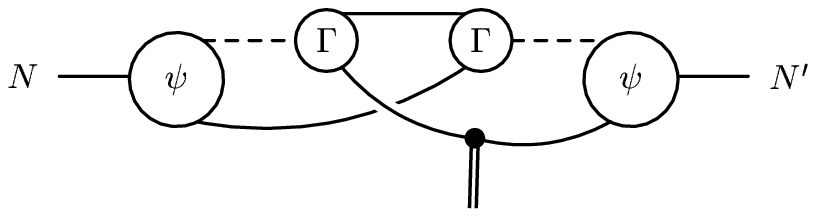,height=1.5cm}}

\vspace*{\baselineskip}

\centering{\ \epsfig{figure=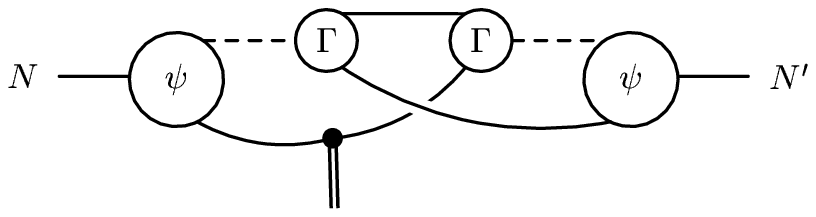,height=1.5cm}}
\caption{Our impulse approximation to the meson-nucleon form factors requires
the calculation of five contributions, which differ in detail for each probe.
$\psi$: $\psi(\ell_1,\ell_2)$ in (\protect\ref{littlepsi}); $\Gamma$:
Bethe-Salpeter-like diquark amplitude in (\protect\ref{gdq}); dashed line:
$\Delta(K)$, diquark propagator in (\protect\ref{dprop}); solid internal
line: $S(q)$, quark propagator in (\protect\ref{qprop}); and double-line:
meson-like probe.  The lowest three diagrams, which describe the interchange
between the dormant quark and the diquark participants, effect the
antisymmetrisation of the nucleon's Fadde'ev amplitude.
\label{diagrams}}
\end{figure}
\begin{eqnarray}
\label{GaPia}{\rm a)}\;\;
f_\pi \tilde E_\pi(k;Q)  &=& B_\pi(k^2)\,, \\
\label{GaPib}{\rm b)}\;\;
f_\pi \tilde E_\pi(k;Q)  &=& \frac{1}{2} \left[B_\pi(k_+^2) +
B_\pi(k_-^2)\right] \,,
\end{eqnarray}
with $k_\pm = k \pm Q/2$.  The first, which assumes no change off shell, has
been used with phenomenological success in a variety of calculations that
explore meson-loop corrections to hadronic observables\cite{piloops}; the
second\cite{cdragw} allows some minimal dependence on $k\cdot Q$, $Q^2$; and
for $Q=0$ both satisfy the constraint of Eq.~(\ref{actualpi})
[cf. Eq.~(\ref{piamp})].

As with the electromagnetic form factors, five distinct diagrams contribute
to the nucleon form factors, which are depicted in Fig.~\ref{diagrams}.  For
the $\pi NN$ coupling these diagrams, enumerated from top to bottom, are
mnemonics for the vertices $\Lambda_\pi^{nj}(q,P)$ given in
Eqs.~(\ref{piNN1}-\ref{piNN5}).  As can be anticipated,
$\Lambda_\pi^{2j}(q,P)\equiv 0$ because of parity conservation; i.e., a
Poincar\'e invariant theory can't admit a three-point
pseudoscalar-scalar-scalar coupling.

The pion-nucleon vertex:
\begin{equation}
\label{nucvtx} \Lambda^{j}_\pi(q,P) = \Lambda^{1j}_\pi(q,P) + 2
\sum_{n=2}^5\,\Lambda^{nj}_\pi(q,P)\,,
\end{equation}
is completely expressed in terms of four independent scalar functions
\begin{equation}
\Lambda_\pi^j(q,P) =   \tau^j \gamma_5 \left[\, i f_1
+ \gamma\cdot q \, f_2 + \gamma\cdot R \, f_3  + \sigma_{\mu\nu} R_\mu q_\nu \,
f_4\right]
\end{equation}
where $f_i= f_i(q^2)$, $R= (P^\prime + P)$ and $q\cdot R = 0$ for nucleon
elastic scattering.  From this we construct the pion-nucleon current
\begin{equation}
\label{Jnucleon} J_\pi^j(P^\prime,P)  =  \bar u(P^\prime)\,
\Lambda_\pi^j(q,P) \,u(P)\,,
\end{equation}
and employing the definition of the nucleon spinors, Eqs.~(\ref{spinors}), we
identify the pion-nucleon coupling in Eq.~(\ref{PiNcur}):
\begin{eqnarray}
g_{\pi N N}(q^2)  &=& f_1 - 2\,M\,f_2 + R^2\,f_4 \,.
\end{eqnarray}

\begin{table}[t]
\caption{Calculated couplings compared with: contemporary meson exchange
model values\protect\cite{harry}, where available; experiment in the case of
$g_A$, $r_A$\protect\cite{gAexpt}; a lattice-QCD result for
$\sigma$\protect\cite{latticesigma}; and for $g_\sigma$, $r_\sigma$, as
discussed in connection with
Eqs.~(\protect\ref{onshellg}-\protect\ref{radsig}).  Also for comparison, the
pion model described in Sec.~\protect\ref{Sect:Model}
yields\protect\cite{mark} $r_\pi=0.56\,$fm.  The labels ``a)'' etc.  identify
the results obtained with: Eqs.~(\protect\ref{GaPia},\protect\ref{GaPib}) for
$\pi NN$; Eqs.~(\protect\ref{GaVa},\protect\ref{GaVb},\protect\ref{GaVc}) for
$VNN$; and Eqs.~(\protect\ref{AVVa},\protect\ref{AVVb}) for the axial-vector
coupling.\label{couplings}}
\[
\begin{array}{l|ccc}
            & {\rm Calc.} & {\rm Estimates} & {\rm Expt.} \\\hline
 g_{\pi NN} & ~14.9 & 13.4 & \\\hline
 \langle r_{\pi NN}^2\rangle^{1/2} & \begin{array}{l}
                  a)\;0.71 \\
                  b)\;0.80 \,
                 \end{array} & 0.93 - 1.06 \,{\rm fm} & \\\hline
 g_{\rho NN}   & \begin{array}{l}
                  a)\;5.92 \\
                  b)\;6.26 \\ 
                  c)\;4.82 
                 \end{array} & 6.4 & \\\hline
 f_{\rho NN}   & \begin{array}{l}
                  a)\;15.4 \\
                  b)\;16.6 \\
                  c)\;12.6
                 \end{array} & 13.0 & \\\hline
 \kappa_{\rho} & \begin{array}{l}
                  a)\;2.57 \\
                  b)\;2.64 \\
                  c)\;2.61
                 \end{array} & 2.0 & \\\hline
 g_{\omega NN} & \begin{array}{l}
                  a)\;9.74 \\
                  b)\;10.2 \\
                  c)\;11.5 
                 \end{array} & 7 - 10.5 & \\\hline
 f_{\omega NN} & \begin{array}{l}
                  a)\;9.62 \\
                  b)\;10.7  \\
                  c)\;4.39
                 \end{array} & & \\\hline
 \kappa_{\omega} & \begin{array}{l}
                  a)\;0.99 \\
                  b)\;1.04\\
                  c)\;0.38
                 \end{array} & & \\\hline
 g_A             & \begin{array}{l}
                  a)\;0.80 \\
                  b)\;0.99
                 \end{array} & & 1.259 \pm 0.017 \\\hline
 \langle r_A^2\rangle^{1/2} & \begin{array}{l}
                  a)\;0.75 \\
                  b)\;0.75\,
                 \end{array} & & 0.68 \pm 0.12 \,{\rm fm} \\\hline
 \sigma/M_N & ~0.015 & 0.019\pm 0.05  & \\\hline
 g_{\sigma} & 9.3 & 10  \\\hline
 \langle r_{\sigma NN}^2\rangle^{1/2} & ~0.89 & 1.2\,{\rm fm} & \\\hline
\end{array}
\]
\end{table}

\begin{figure}[t]
\centering{\epsfig{figure=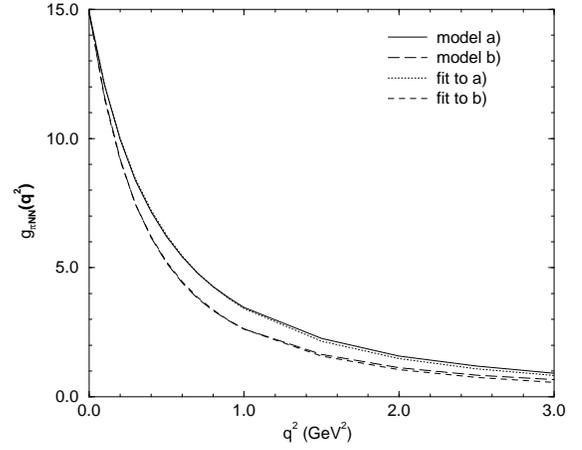,height=6.0cm}} \vspace*{\baselineskip}
\caption{Calculated pion-nucleon form factor, compared with dipole fits,
Eqs.~(\protect\ref{dipolepiNN},\protect\ref{polemasses}).\label{piNNFF}}
\end{figure}

Using Monte-Carlo methods to evaluate the integrals we obtain the coupling,
$g_{\pi NN}:= g_{\pi NN}(q^2=0)$, in Table~\ref{couplings}.  It is 11\% too
large.  (Our statistical error is always $< 1\,$\%.)  We anticipate that
retaining pseudovector components in $\Gamma_\pi^j(k;Q)$ and an axial-vector
diquark correlation will only slightly affect this value as long as they are
constrained {\it consistently} with the model.  An {\it ad hoc} addition of
the pseudovector components can have large effects\cite{peterparis}.

The relative strength of the contribution from each diagram in
Fig.~\ref{diagrams} is presented in Table~\ref{tablerelative}.  We observe
that the diquark breakup terms are just as important here as they were in the
calculation of the nucleon charge radii and magnetic moments\cite{jacques}.
These diagrams are the true measure of the diquark's composite nature, which
is not captured by simply adding a diquark ``vertex function.''

\begin{table}[t]
\caption{Relative contribution to the couplings of each of the terms
represented mnemonically by the five diagrams in Fig.~\protect\ref{diagrams}.
In building this table we used the amplitudes in Eq.~(\protect\ref{GaPia})
for the $\pi$, Eqs.~(\protect\ref{GaVa},\protect\ref{GaVc}) for the vector
couplings, and the dressed-quark-axial-vector vertex of
Eqs.~(\protect\ref{AVAnsatz},\protect\ref{AVVb}).  In all cases the crucial
role of the diquark breakup diagrams is evident.  Diagram 2 will contribute
to all processes if an axial-vector diquark correlation is included.
\label{tablerelative}}
\[
\begin{array}{lr|rrrrr}
& {\rm diagram} & 1 & 2 & 3 & 4 & 5 \\\hline
g_{\pi NN}  && 0.65 & 0.00  & 0.07 & 0.14 & 0.14\\
g_{\rho NN} & {\rm a)} & 0.74 & 0.00 &-0.06  & 0.16 & 0.16 \\
            & {\rm c)} & 0.73 & 0.00 &-0.11  & 0.19 & 0.19 \\
f_{\rho NN} & {\rm a)} & 0.64 & 0.00 & 0.10  & 0.13 & 0.13 \\
            & {\rm c)} & 0.64 & 0.00 & 0.12  & 0.12 & 0.12 \\
g_{\omega NN}&{\rm a)} & 0.45 & 0.31 & 0.04 & 0.10 & 0.10\\
             &{\rm c)} & 0.31 & 0.49 & 0.05 & 0.08 & 0.08\\
f_{\omega NN}&{\rm a)} & 1.04 & -0.28 & -0.16 & 0.20 & 0.20 \\
             &{\rm c)} & 1.81 & -1.16 & -0.35 & 0.35 & 0.35 \\
g_A          && 0.63 & 0.00 & 0.09 & 0.14 & 0.14 \\
\sigma       && 0.58 & 0.19 & 0.03 & 0.10 & 0.10
\end{array}
\]
\end{table}
\hspace*{-\parindent}

The $\pi NN$ form factor calculated using both off-shell {\it Ans\"atze},
Eqs.~(\ref{GaPia},\ref{GaPib}), is plotted in Fig.~\ref{piNNFF}.  We have
performed monopole and dipole fits to our calculated result:
\begin{equation}
\label{dipolepiNN}
g_{\pi N N}(q^2)= \frac{g_{\pi N N}}{(1+q^2/\Lambda_\pi^2)^n}\,,\; n=1,2\,,
\end{equation}
and obtain pole masses, in GeV,
\begin{equation}
\label{polemasses}
\begin{array}{l|cc}
        & n=1 & n=2 \\\hline
{\rm Eq.~(\protect\ref{GaPia})} & 0.63 & 0.96 \\
{\rm Eq.~(\protect\ref{GaPib})} & 0.57 & 0.85 
\end{array}\,.
\end{equation}
The dipole form provides an accurate interpolation on the entire range shown.
However, the monopole form is only accurate for $q^2\lsim 0.4\,$GeV$^2$,
overestimating the result by $\sim 70\,$\% at $q^2=3.0\,$GeV$^2$.  (Requiring
that the fits are accurate in the neighbourhood of $q^2=0$ ensures
$\Lambda_\pi^{\rm dipole}/\Lambda_\pi^{\rm monopole} \approx \surd 2$.)  Thus
our calculations favour soft form factors, in semi-quantitative agreement
with those employed in Ref.\cite{harry} and advocated in Ref.\cite{tony}.  We
can further quantify this by introducing a pionic radius of the nucleon:
\begin{equation}
\langle r_{\pi NN}^2 \rangle := -\frac{6}{g_{\pi NN}}\, 
\left.\frac{d g_{\pi NN}(q^2)}{dq^2}\right|_{q^2=0}\,.
\end{equation}
Our calculated value is presented in Table~\ref{couplings} and can be
compared with the analogous tabulated quantities.  It is almost three-times
larger than $r_{\pi NN} \sim 0.3\,$fm inferred from Ref.~\cite{machleidt}.

\section{Vector-meson Nucleon Coupling}
\label{secVN}
In this section we consider $\omega$-$NN$- and $\rho$-$NN$-like interactions;
i.e., isoscalar-vector and isovector-vector couplings.  The
vector-meson--nucleon current is
\begin{eqnarray}
\lefteqn{J_\mu^{V\alpha}(P^\prime,P) = }\\ && \nonumber i\,\bar
u(P^\prime)\frac{\tau^\alpha}{2}
\left(\gamma_\mu \, F_1^V(q^2) + \frac{1}{2 M}\, \sigma_{\mu\nu}\,
q_\nu\, F_2^V(q^2)\right)u(P)\,,
\end{eqnarray}
with $\tau^0:= {\rm diag}(1,1)$ and $\tau^{1,2,3}$ the usual Pauli matrices.
Although the complete specification of a fermion-vector boson vertex:
\begin{equation}
\Lambda_\mu^\alpha(q,P):= \frac{\tau^\alpha}{2}\,\Lambda_\mu(q,P)\,,
\end{equation}
requires twelve independent scalar functions,
\begin{eqnarray}
\nonumber 
\lefteqn{i\Lambda_\mu(q,P)  =  i \gamma_\mu\,f_1
+ i\sigma_{\mu\nu}\,q_\nu\,f_2
+ R_\mu\,f_3 
+ i\gamma\cdot R\,R_\mu\,f_4
}\\
&& 
+ i\sigma_{\nu\rho}\,R_\mu\,q_\nu\,R_\rho\,f_5
+ i\gamma_5\gamma_\nu\,\varepsilon_{\mu\nu\rho\sigma}\,
         q_\rho\,R_\sigma\,f_6\,+ \ldots\,,
\end{eqnarray}
using Eq.~(\ref{spinors}) only the six shown explicitly contribute to
$F_{1,2}$:
\begin{eqnarray}
\label{F1def}
\lefteqn{F_1  = }\\
&& \nonumber f_1 
+ 2\,M\,f_3 - 4\, M^2\,f_4 - 2\,M\,q^2\,f_{5} - q^2\,f_{6}\,,\\
\label{F2def}
\lefteqn{F_2  = }\\
&& \nonumber  2\,M\,f_2
-2\,M\,f_3 + 4\,M^2\,f_4 + 2\,M\,f_{5} - 4\,M^2\,f_{6}\,.
\end{eqnarray}
The coupling strengths relevant for comparison with potential models are
\begin{equation}
g_{V NN} := F_1^V(0)\,,\;
f_{V NN} := F_2^V(0)\,,\;\kappa_V:=\frac{f_{VNN}}{g_{VNN}}
\end{equation}
because these are $t$-channel elastic scattering models.  However, we note
that $q^2=0$ is a far off-shell point for the $\omega$ and $\rho$ of the
strong interaction spectrum, for which $(-q^2)= M_V^2 \approx 0.6\,$GeV$^2$.
Hence a calculation of these coupling constants is only possible after an
off-shell extrapolation of the vector meson Bethe-Salpeter amplitude is
specified.

Quantitatively reliable numerical solutions of the vector meson
Bethe-Salpeter equation have recently become available\cite{marisvector},
however, an algebraic {\it Ansatz} compatible with our parametrisation of the
dressed-quark propagator, Eqs.(\ref{ssm},\ref{svm}), is not yet available.
Hence, we use {\it Ans\"atze} motivated by an extensive study of light- and
heavy-meson observables\cite{misha}:
\begin{equation}
\label{BSAV}
\Gamma_\mu^\alpha(k;Q) = \frac{1}{{\cal N}_V} \left(\gamma_\mu - \frac{Q_\mu
\gamma\cdot Q}{Q^2}\right) \varphi(k^2) \,\tau^\alpha \,,
\end{equation}
\begin{eqnarray}
\label{GaVa}{\rm a)}\;\;
\varphi(k^2) &=& 1/(1+k^4/\omega^4)\,, \\
\label{GaVb}{\rm b)}\;\;
\varphi(k^2) &=& \left[{\cal F}(k^2/\omega^2)\right]^2\,,
\end{eqnarray}
with $\omega=0.515\,$GeV and the normalisation: ${\cal N}_V$, determined
canonically, Eq.~(\ref{Vnorm}).  With these simple forms of
$\Gamma_\mu^\alpha(k;Q)$ we obtain the following values of the
electromagnetic and strong coupling constants:
\begin{equation}
\label{vectorcouplings}
\begin{array}{l|cccc}
       & a)     & b)    & c)    & {\rm Obs.\cite{pdg}} \\\hline
g_\rho & 6.57 &  6.05 & & 5.03 \pm 0.012 \\
g_{\rho\pi\pi} & 8.75 & 10.7 & 8.52 & 6.05 \pm 0.02
\end{array}\,,
\end{equation}
results which suggest that errors of up-to 40\% could arise in nucleon
calculations involving these amplitudes.

The calculation of the vector-meson--nucleon current is now straightforward
with $F_{1,2}$ determined by calculating the integrals in
Eqs.~(\ref{V1}-\ref{V5}) and combining their contributions according to
Eqs.~(\ref{F1def},\ref{F2def},\ref{fullvector}).  In this way we obtain the
couplings presented in Table~\ref{couplings}, with the relative strength of
the contribution from each diagram given in Table~\ref{tablerelative}.

The couplings are in semi-quantitative agreement with those inferred from
meson-exchange models\cite{harry} {\it except} in the case of $f_{\omega
NN}$.  Using Eqs.~(\ref{GaVa},\ref{GaVb}) we obtain $f_{\omega NN} \approx
g_{\omega NN}$ while contemporary phenomenological models, which are only
weakly sensitive to $f_{\omega NN}$, assume it to be zero.  To determine the
extent to which this result is model dependent we also calculate the
couplings using
\begin{figure}
\centering{\ \epsfig{figure=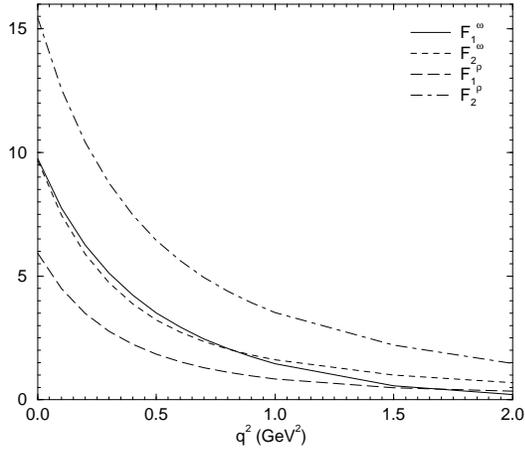,height=6.0cm}} 
\caption{Calculated vector-meson--nucleon form factors.  A good interpolation
of the results is obtained with
Eqs.~(\protect\ref{polefit},\protect\ref{polemassesII}).\label{VecMes}}
\end{figure}
\begin{equation}
\label{GaVc}
{\rm c)} \;\; \varphi(k^2)= \frac{1}{{\cal N}_V}\,B_V(k^2)\,,
\end{equation}
where $B_V(k^2)$ is obtained from Eqs.~(\ref{Sinv}-\ref{svm})
with\cite{hawes}  $\bar m \to 0$ and 
\begin{equation}
b_0 \to b_0^V = 0.044\,,\;
b_1 \to b_1^V = 0.580\,,\;
b_3 \to b_3^V = 0.462\,.
\end{equation}
The amplitude is canonically normalised via Eq.~(\ref{Vnorm}) and yields
$g_{\rho\pi\pi}$ in Eq.~(\ref{vectorcouplings}).  

The nucleon couplings are given in Table~\ref{couplings} with the relative
strength of the contribution from each diagram presented in
Table~\ref{tablerelative}.  In this case, while the other couplings change by
$\lsim 20\,$\%, we find $ f_{\omega^{\rm c)} NN} \approx 0.4\,g_{\omega^{\rm
c)} NN}$.  The reason appears in Table~\ref{tablerelative}: the strength of
diagram 2 is much increased, and while it doesn't contribute at all for the
$\rho$ and is additive for $g_{\omega NN}$, it is a destructive contribution
to $f_{\omega NN}$.  This sensitivity to cancellations involving diagram 2
repeats the pattern we observed in calculating the nucleon's isoscalar
electromagnetic form factor\cite{jacques} and hence $f_{\omega NN}$ is
sensitive to the omission of the axial-vector diquark correlation.

The vector-meson--nucleon form factors calculated using Eq.~(\ref{GaVa}) are
depicted in Fig.~\ref{VecMes} and the quadrupole
\begin{equation}
\label{polefit}
F_n^V(q^2) = F_n^V(0)\,\frac{1}{(1+q^2/\Lambda^V_n)^3}
\end{equation}
with pole masses, in GeV,
\begin{equation}
\label{polemassesII}
\begin{array}{cccc}
\Lambda_1^\omega & \Lambda_2^\omega & \Lambda_1^\rho & \Lambda_2^\rho \\
1.12    & 1.11 &  1.06 & 1.18      
\end{array}
\end{equation}
provides an excellent interpolation of the results.  Again, these are soft
form factors.  

On the domain explored, our results for the $VNN$ form factor are
qualitatively unaffected by employing a monopole for $\varphi(k^2)$ in
Eq.~(\ref{BSAV}).

\section{Axial-vector Nucleon Coupling}
\label{secgA}
Neutron $\beta$-decay is described by the axial-vector--nucleon current
\begin{eqnarray}
\label{AVcur}
\lefteqn{ J_{5\mu}^j(P',P)  =  i \bar u(P')\, \Lambda_{5\mu}^j(q,P) u(P)}\\
& =: & i \bar u(P')\, \gamma_5\frac{\tau^j}{2} \left[\gamma_\mu g_A(q^2) +
q_\mu g_P(q^2) \right]u(P) \,,
\end{eqnarray}
which involves two form factors: $g_A(q^2)$ is the axial-vector form factor
of the nucleon and $g_P(q^2)$ is the induced pseudoscalar form factor.  The
complete specification of a fermion-axial-vector vertex:
\begin{equation}
\Lambda_{5\mu}^j(q,P):= \frac{\tau^j}{2}\,\Lambda_{5\mu}(q,P)\,,
\end{equation}
requires twelve independent scalar functions, 
\begin{eqnarray}
\nonumber \lefteqn{\Lambda_{5\mu}(q,P) = }\\ & & \gamma_5\gamma_\mu\, f_1 +
\gamma_5\sigma_{\mu\nu} R_\nu\,f_2 + \epsilon_{\mu\nu\rho\sigma}\,\gamma_\nu
q_\rho R_\sigma\,f_3 + \ldots
\end{eqnarray}
but using Eqs.~(\ref{spinors}) only the three shown explicitly contribute to
the axial-vector form factor:
\begin{equation}
\label{ga123}
g_A(q^2) = f_1 - 2 M f_2 - q^2 f_3 \,.
\end{equation}

In the chiral limit and in the neighbourhood of $q^2=0$\cite{mrt98,mr97}
\begin{equation}
\Lambda_{5\mu}^j(q,P) = {\it regular} 
+ \frac{q_\mu}{q^2}\,f_\pi\,\Lambda_\pi^j(q,P)\,,
\end{equation}
where $\Lambda_\pi^j(q,P)$ is the pion-fermion vertex and {\it regular}
denotes non-pole terms.  It follows that in this neighbourhood the induced
pseudoscalar coupling is dominant and, using Eq.~(\ref{PiNcur}), is
determined by the pion-nucleon coupling:
\begin{eqnarray}
\lefteqn{\left.q^2 J_{5\mu}^j(P^\prime,P)\right|_{q^2=0}}
\nonumber \\
& = &  
\label{gPgpiNN}
q_\mu\,f_\pi\,g_{\pi NN}(q^2=0)\,
\bar u(P^\prime) i\tau^j\gamma_5 \,u(P)\\
& = & q_\mu\,\bar u(P^\prime) \, 
\gamma_5 \frac{\tau^j}{2}\,[q^2 g_P(q^2)]_{q^2=0}\,u(P)\,.
\end{eqnarray}
Current conservation: $q_\mu\,J_{5\mu}^j(q,P)=0$, which using
Eqs.~(\ref{spinors}) entails
\begin{equation}
-\,2 M g_A(q^2=0) + \left.\left[q^2\,g_P(q^2)\right]\right|_{q^2=0} = 0\,,
\end{equation}
then yields the Goldberger-Treiman relation:
\begin{equation}
\label{GTR}
M\,g_A(q^2=0) = f_\pi\,g_{\pi NN}(q^2=0)\,.
\end{equation}
This brief analysis emphasises that $g_A(q^2)$ is the {\it regular} part of
the axial-vector--nucleon current.

To calculate $g_A(q^2)$ in impulse approximation we must specify the
dressed-quark-axial-vector vertex: $\Gamma_{5\mu}^j(k;Q)$.  It satisfies a
Ward-Takahashi identity, which in the chiral limit is
\begin{eqnarray}
-i Q_\mu \Gamma_{5\mu}^j(k;Q)  & = & 
S^{-1}(k_+)\gamma_5\frac{\tau^j}{2}
+  \gamma_5\frac{\tau^j}{2} S^{-1}(k_-) \,.
\end{eqnarray}
This identity is solved by
\begin{eqnarray}
\label{AVTotal}
& & \Gamma_{5\mu}^j(k;Q) = 
\Gamma_{5\mu}^{Rj}(k;Q)  
+ \left[i \frac{Q_\mu}{Q^2}\,2\,\Sigma_B(k_+^2,k_-^2)\right] \,,\\
\nonumber 
\lefteqn{\Gamma_{5\mu}^{Rj}(k;Q) = \Gamma_{5\mu}^{Tj}(k;Q)}\\
&+ & \label{AVAnsatz}
 \gamma_5\frac{\tau^j}{2}
\left(\rule{0mm}{1.5em}
\gamma_\mu\,\Sigma_A(k_+^2,k_-^2)
+ 2\,k_\mu\,\gamma\cdot k\,\Delta_A(k_+^2,k_-^2)
\right)  \,,
\end{eqnarray}
where $Q_\mu\Gamma_{5\mu}^{Tj}(k;Q) = 0$ but $\Gamma_{5\mu}^{Tj}(k;Q)$ is
otherwise unconstrained by the Ward-Takahashi identity and
\begin{eqnarray}
\Sigma_f(p^2,q^2) &:= & \case{1}{2}\,[f(p^2)+f(q^2)]\,,\\
\Delta_f(p^2,q^2) & := & \frac{f(p^2)-f(q^2)}{p^2-q^2}\,.
\end{eqnarray}

The parenthesised term in Eq.~(\ref{AVTotal}) makes explicit the simple
kinematic singularity associated with the pion pole.\footnote{\label{ft1}
Using $(k_\mu/k\cdot Q)\,\Sigma_B(k_+^2,k_-^2)$ in Eq.~(\ref{AVTotal})
instead of the parenthesised term is inadequate in this respect.  Further, to
exacerbate this flaw, it also introduces nonintegrable singularities in
diagrams $3$-$5$.}
It is directly connected with the nucleon's induced pseudoscalar form factor
and, using Eqs.~(\ref{piamp},\ref{GaPib}), clearly saturates
Eq.~(\ref{gPgpiNN}).  The regular part of the vertex, Eq.~(\ref{AVAnsatz}),
is primarily responsible for the nucleon's axial-vector form factor and in
our calculations we complete its definition using either of two {\it
Ans\"atze} for the transverse part:
\begin{eqnarray}
\label{AVVa}
{\rm a})\;\; \Gamma_{5\mu}^{Tj}(k;Q) & = & 0\,,\\
\label{AVVb}
{\rm b})\;\; 
\Gamma_{5\mu}^{Tj}(k;Q)  & = &  \case{1}{\sqrt{2}}\,
\frac{f_{a_1} m_{a_1}}{Q^2 + m_{a_1}^2}\,
\Gamma_\mu^{a_1 j}(k;Q)\,,
\end{eqnarray}
where $\Gamma_\mu^{a_1 j}(k;Q)$ is the $a_1$-meson Bethe-Salpeter amplitude,
which is given explicitly in Eq.~(\ref{BSAa1}).  Model $a)$:
Eqs.~(\ref{AVAnsatz},\ref{AVVa}), is a minimal {\it Ansatz} that correctly
isolates the pion pole.  Model $b)$: Eqs.~(\ref{AVAnsatz},\ref{AVVb}), is
kindred to that advocated for the dressed-quark-vector vertex in
Ref.~\cite{marispion}.  It recognises that the dressed-quark-axial-vector
vertex has a pole at $Q^2=-m_{a_1}^2$ with residue $f_{a_1} m_{a_1}$ where
$f_{a_1}$ is the weak decay constant, and implements a model to represent the
off-shell remnant of this contribution.

$g_A(q^2)$ is obtained by evaluating the integrals in
Eqs.~(\ref{AV1}-\ref{AV5}) and combining their contributions according to
Eqs.~(\ref{ga123},\ref{fullaxial}).  The calculated coupling is presented in
Table~\ref{couplings} with the relative strength of the contribution from
each diagram given in Table~\ref{tablerelative}.\footnote{We are unable to
reproduce the large value of $g_A$ obtained in Ref.~\protect\cite{reinhard}.
Some of the discrepancy may be due to our simplified representation of the
quark$+$scalar-diquark nucleon spinor in
Eq.~(\protect\ref{Psi})\protect\cite{rapriv}.  However, that does not
diminish the importance of $\Gamma_{5\mu}^{Tj}$.}
\begin{figure}
\centering{\ \epsfig{figure=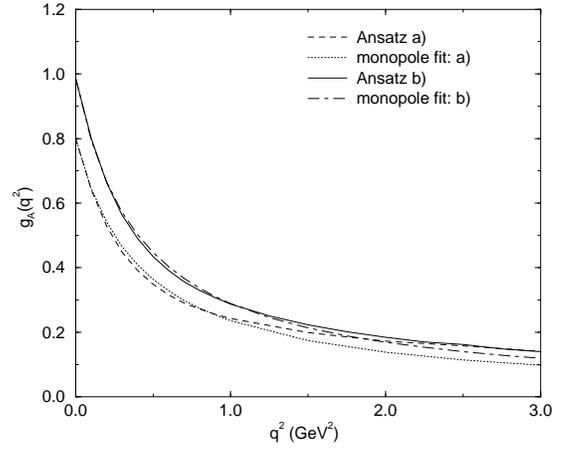,height=6.0cm}} 
\caption{Axial-vector--nucleon form factor calculated using the {\it
Ans\"atze} of Eqs.~(\protect\ref{AVVa},\protect\ref{AVVb}) compared with a
monopole fit, Eqs.~(\protect\ref{gAmono},\protect\ref{gAmasses}). \label{gA}}
\end{figure}

The axial-vector form factor is depicted in Fig.~\ref{gA}.  It is important
and interesting to note that the dominant, orbital $e_2^{a_1}$-term in
$\Gamma_\mu^{a_1}$ contributes $< 10\,$\% to $g_A(q^2)$ on the illustrated
domain, increasing from $0$\% with increasing $q^2$; i.e., the bulk of the
difference between the $a)$ and $b)$ calculations arises from the
$e_1^{a_1}$-term.  In Fig.~\ref{gA} we also plot a monopole fit to each
calculation
\begin{equation}
\label{gAmono}
g_A(q^2)= g_A(0)\,\frac{1}{(1+q^2/\Lambda_A^2)}
\end{equation}
with pole masses, in GeV, 
\begin{equation}
\label{gAmasses}
\begin{array}{l|cc}
          & {\rm a}) & {\rm b}) \\\hline
\Lambda_A & 0.65 & 0.64
\end{array}
\end{equation}
The fit provides a better representation for {\it Ansatz} $b)$ than for $a)$,
and we judge $b)$ to be the more realistic model.  The axial radius of the
nucleon is
\begin{equation}
\langle r_A^2 \rangle := -\frac{6}{g_A(0)}\, 
\left.\frac{d g_A(q^2)}{dq^2}\right|_{q^2=0}
\end{equation}
and our calculated value is presented in Table~\ref{couplings}: $r_{A}
\approx r_{\pi NN}$ in accordance with Ref.~\cite{tony}.

Even though our framework manifestly preserves the axial-vector
Ward-Takahashi identity the model is not guaranteed to satisfy the
Goldberger-Treiman relation because we employ {\it Ans\"atze} for the
Fadde'ev amplitude and dressed-quark-axial-vector vertex that need not be
mutually consistent.  For example, in deriving Eq.~(\ref{GTR}) we used
Eqs.~(\ref{spinors}), which introduce $M$, however, $\Psi$ in Eq.~(\ref{Psi})
is not the solution of a Fadde'ev equation with eigenvalue $M$.  Indeed,
without an axial-vector diquark correlation, the calculated nucleon mass is
30-50\% too large\cite{ishii,raAB,regFB}.  The following comparisons (in GeV)
exhibit this uncertainty:
\begin{equation}
\left.
\begin{array}{l}
{\rm a})\;\; M g_A(0) = 0.75 - 1.1 \\
{\rm b})\;\; M g_A(0) = 0.93 - 1.4 
\end{array}
\right\}\;\;{\rm cf.}\;\;  f_\pi g_{\pi N N} = 1.37\,.
\end{equation}
and show that model {\it Ansatz} $b)$ is broadly consistent with the
Goldberger-Treiman relation.  

A comparison between the results obtained with the two vertex {\it Ans\"atze}
demonstrates that a calculation of the dressed-quark-axial-vector vertex,
akin to that of the dressed-quark-vector vertex in Ref.~\cite{marispion},
would be very helpful in demarcating the importance of axial-vector diquark
correlations.  As shown by the model $b)$ calculation, $\Gamma_{5\mu}^{Tj}$
easily provides contributions of the same order of magnitude as that which
might be anticipated from an axial-vector diquark.

\section{Nucleon Sigma Term}
\label{secsig}
As a final application we explore the nucleon $\sigma$-term:
\begin{equation}\label{sigma}
\sigma(q^2)\, \bar u(P') u(P):= \langle P'|\, m (\bar u u + \bar d
d)\,|P\rangle \,,
\end{equation}
$\sigma:= \sigma(q^2=0)$, which is the in-nucleon expectation value of the
explicit chiral symmetry breaking term in the QCD Lagrangian.  The general
form for a fermion-scalar vertex is
\begin{equation}
\Lambda_{\mbox{\large\boldmath $1$}}(q,P) = f_1 + i \gamma\cdot q \,f_2 + i
\gamma\cdot R \,f_3 + i \sigma_{\mu\nu} R_\mu q_\nu \,f_4 \,,
\end{equation}
however, using Eqs.~(\ref{spinors}) we find
\begin{eqnarray}
J_{\mbox{\large\boldmath $1$}}(P',P) 
& :=& \bar u(P') \Lambda_{\mbox{\large\boldmath $1$}}(q,P) u(P) \\
& = & s(q^2) \,\bar u(P') \,u(P) \,,\\
\label{sigmaadd}
s(q^2) & = & f_1 - 2 M f_3 + q^2 f_4 \,.
\end{eqnarray}

To evaluate the matrix element in Eq.~(\ref{sigma}) we need the
dressed-quark-scalar vertex, which is an analogue of the
dressed-quark-axial-vector vertex used in Sec.~\ref{secgA}.  In this case,
however, there isn't a Ward-Takahashi identity to help us.  Instead, we
calculate the vertex by solving an inhomogeneous Bethe-Salpeter equation using
a simple, separable model for the quark-antiquark scattering
kernel\cite{bsesep} that has been used successfully in a variety of
phenomenological applications\cite{a1b1,exotics}.  

The inhomogeneous vertex equation in the separable model is
\begin{eqnarray}\label{IBS}
\lefteqn{\Gamma_{\mbox{\large\boldmath $1$}}(k;Q) =} \\ && \nonumber
\mbox{\large\boldmath $1$} - \case{4}{3} \int \frac{d^4
q}{(2\pi)^4} \Delta(k-q) \gamma_\mu S(q_+) 
\Gamma_{\mbox{\large\boldmath $1$}}(q;Q) S(q_-) \gamma_\mu \,,
\end{eqnarray}
with the interaction
\begin{equation}\label{SepAn}
\Delta(k-q) = G(k^2)G(q^2) + k\cdot q \, F(k^2) G(k^2)\,,
\end{equation}
where
\begin{equation}\label{FG}
F(k^2) = \frac{1}{a}[A(k^2)-1] \hspace{5mm} G(k^2) =
\frac{1}{b}[B(k^2)-\tilde{m}]\,,
\end{equation}
$a = \bar a \lambda^2$, $b = \bar b \lambda^2$, and $A(k^2)$, $B(k^2)$ are
obtained in the usual way from Eqs.~(\ref{ssa1},\ref{sva1}) with $\tilde m =
\hat m \lambda$ and $b_2^{a_1}\to b_2$.  The separable model was constrained
to fit $\pi$ and $K$ properties, as discussed in detail in
Ref.~\cite{bsesep}.

Using this model the most general form of the scalar vertex is
\begin{equation}
\label{Ga1}
\Gamma_{\mbox{\large\boldmath $1$}}(k;Q) = \mbox{\large\boldmath $1$}\,
g_1 + i k\cdot Q\gamma\cdot Q \, g_2 + i \gamma\cdot k\,g_3 \,,
\end{equation}
where $g_i= g_i(k;Q)$.  Substituting Eq.~(\ref{Ga1}) in Eq.~(\ref{IBS}) we
obtain the solution
\begin{eqnarray}\label{Ga1Sol}
\lefteqn{
\Gamma_{\mbox{\large\boldmath $1$}}(k;Q) = \mbox{\large\boldmath $1$}  +
t_1(Q^2)\, G(k^2)} \\ 
&& \nonumber
+ \, i\,t_2(Q^2) \,F(k^2)\frac{k\cdot Q \,\gamma\cdot Q}{Q^2}
+ \,i\,t_3(Q^2) F(k^2)  \gamma\cdot k \,, \nonumber
\end{eqnarray}
with $t_i(Q^2)$ determined functions of their argument.

The $\sigma$-term is only sensitive to the vertex at $Q^2=0$, where the
explicit form of the solution reduces to
\begin{equation}\label{Ga1Sol0}
\Gamma_{\mbox{\large\boldmath $1$}}(k;Q)|_{Q^2=0} = \mbox{\large\boldmath
$1$}  + t_1(0)\, G(k^2) + t_3(0)\, F(k^2)\, i \gamma\cdot k \,,
\end{equation}
with $t_1(0)=0.242\,$GeV, $t_3(0)=-0.0140\,$GeV.  It is important to note
from Eqs.~(\ref{FG},\ref{Ga1Sol0}) that the $t_1$-term contributes $1.4
\times(B(k^2)-\tilde{m})/\sqrt{b}$ so that at $k^2=0$ it is 6-times larger
than the bare term; i.e., it is dominant in the infrared.  That is to be
expected because it represents the effect of the nonperturbative dynamical
chiral symmetry breaking mechanism in the solution.  This and the other
$g_i$-terms vanish as $k^2\to \infty$, which is a manifestation of asymptotic
freedom in the separable model.

The vertex equation has a solution for all $Q^2$, and that solution exhibits
a pole at the $\sigma$-meson mass; i.e., in the neighbourhood of $(-Q^2)=
m_\sigma^2= (0.715\,{\rm GeV})^2$
\begin{equation}
\label{scalarvtx}
\Gamma_{\mbox{\large\boldmath $1$}}(k;Q) = {\it regular}\; + 
\frac{n_\sigma m_\sigma^2}{Q^2+m_\sigma^2}\,\Gamma_\sigma(k;Q)\,,
\end{equation}
where {\it regular} indicates terms that are regular in this neighbourhood
and $\Gamma_\sigma(k;Q)$ is the canonically normalised $\sigma$-meson
Bethe-Salpeter amplitude, whose form is exactly that of
$(\Gamma_{\mbox{\large\boldmath $1$}}(k;Q) - \mbox{\large\boldmath $1$})$ in
Eq.~(\ref{Ga1Sol}).  The simple pole appears in the functions $t_i(Q^2)$ and
performing a pole fit we find
\begin{equation}
m \,n_\sigma = 3.3\,{\rm MeV}\,.
\end{equation}
$n_\sigma m_\sigma^2$ is the analogue of the residue of the pion pole in the
pseudoscalar vertex\protect\cite{mrt98,mr97}: $-\langle \bar q
q\rangle_\pi/f_\pi$, and its flow under the renormalisation group is
identical.  $m\, n_\sigma$ is renormalisation point independent and its value
can be compared with 
\begin{equation}
\frac{-m \langle \bar q q\rangle_\pi}{f_\pi}\,\frac{1}{m_\sigma^2} =
3.6\,{\rm MeV}.
\end{equation}
We can also define a $\sigma \bar q q$ coupling: 
\begin{equation}
g_{\sigma  \bar q q}:=
\left.\Gamma_\sigma(0;Q)\right|_{Q^2=-m_\sigma^2} = 12.6\,,
\end{equation}
whose magnitude can be placed in context via a comparison with $g_{\pi \bar q
q}=11.8$ obtained using the separable model's analogue of the quark-level
Goldberger-Treiman relation, Eq.~(\ref{piamp}).

To calculate the expectation value in Eq.~(\ref{sigma}) we use
\begin{equation}
\label{sigmavtx}
\Gamma_{m}(k;Q) = m \,\Gamma_{\mbox{\large\boldmath $1$}}(k;Q)
\end{equation}
as our impulse approximation probe in Eqs.~(\ref{S1}-\ref{S5}) and obtain
$\sigma(q^2)=s(q^2)$ by combining the contributions according to
Eqs.~(\ref{sigmaadd},\ref{fullscalar}).  This yields the value of $\sigma$
presented in Table~\ref{couplings}, with the relative strength of the
contribution from the various diagrams listed in Table~\ref{tablerelative}.

\begin{figure}[t]
\centering{\ \epsfig{figure=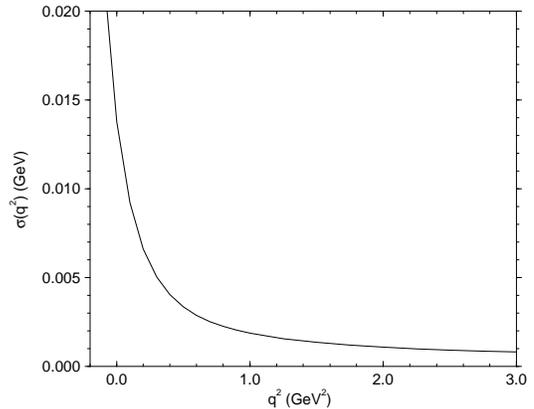,height=5.5cm}} \vspace*{\baselineskip}
\caption{Our calculated $\sigma(q^2)$.  The rapid increase with decreasing $q^2$
is associated with the evolution to the $\sigma$-meson pole.  On this scale,
$\sigma(q^2)$ calculated without the $t_{2,3}(Q^2)$ contributions is
indistinguishable from the full calculation. \label{sigFF}}
\end{figure}

The form factor is depicted in Fig.~\ref{sigFF} where the evolution to the
$\sigma$-meson pole is evident.  Fitting ($t = -q^2$)
\begin{equation}
\label{atpole}
\sigma(t) = g_{\sigma NN}\,\frac{m \,n_\sigma}{1-t/m_\sigma^2}\,,\;
t\in [0.1,0.5]\,{\rm GeV}^2\,,
\end{equation}
which isolates the residue associated with $\Gamma_m(k;Q)$, we obtain the
on-shell coupling: $g_{\sigma NN} = 27.3$.  This coupling can also be
calculated directly using the solution of the homogeneous Bethe-Salpeter
equation and that yields
\begin{equation}
g_{\sigma NN} = 27.7\,,
\end{equation}
in agreement within Monte-Carlo errors.  Equation~(\ref{atpole}) alone
overestimates the magnitude of our calculated $\sigma(t)$ everywhere except
in the neighbourhood of the pole.   

As the lowest-mass pole-solution of Eq.~(\ref{IBS}), our $\sigma$-meson is
distinct from the phenomenological meson introduced in potential models to
mock-up two-pion exchange.\footnote{The separable model\protect\cite{bsesep}
realises a rainbow-ladder truncation of the quark DSE and meson
Bethe-Salpeter equation, which is likely inaccurate in the $0^{++}$
channel\protect\cite{cdrqcII}.  The defect is tied to the difficulties
encountered in understanding the composition of scalar resonances below
$1.4\,$GeV\protect\cite{mikescalars}.}
However, we can estimate a coupling relevant to meson exchange models by
introducing $g_\sigma(t)$:
\begin{equation}
\label{onshellg}
\sigma(t) =: g_{\sigma}(t)\,\frac{m\,n_\sigma}{1-t/m_\sigma^2}\,,
\end{equation}
and a fit to our calculated $\sigma(t)$ yields
\begin{equation}
g_{\sigma}(t) = 1.61 + 
2.61\,\frac{1}{(1-t/\Lambda_\sigma^2)^{10}}\,,\;
\Lambda_\sigma=1.56\,{\rm GeV}\,,
\end{equation}
where the large exponent merely reflects the rapid
evolution from bound state to continuum dominance of the vertex in the
spacelike region.  At the mock-$\sigma$-mass: $m_\sigma^{2\pi}=0.5\,$GeV,
\begin{equation}
g_{\sigma}:= g_{\sigma}((m_\sigma^{2\pi})^2) = 9.3\,,
\end{equation}
which is listed in Table~\ref{couplings} and compared with a
phenomenologically inferred value: $g_\sigma = 10$\cite{Friman}.  We note 
\hspace*{-\parindent}that $g_{\sigma}(4\,m_\pi^2) = 5.2$ so that this
comparison is meaningful on a relevant phenomenological domain.  Further,
$g_\sigma(q^2\to\infty)=1.61$ and we therefore find that $\sigma(q^2)$ is
well approximated by a single monopole for $q^2>1\,$GeV$^2$.  However, the
residue is very different from the on-shell value.  The scalar radius of the
nucleon is obtained from
\begin{equation}
\label{radsig}
\langle r_{\sigma NN}^2 \rangle := -\frac{6}{\sigma}\, 
\left.\frac{d \sigma(q^2)}{dq^2}\right|_{q^2=0}
\end{equation}
and our calculated result is listed in Table~\ref{couplings}, in comparison
with an inferred value\cite{Friman}.

\section{Summary and Conclusion}
\label{secSC}
In Ref.~\cite{jacques} we introduced a simple model of the nucleon's Fadde'ev
amplitude that represents the nucleon as a bound state whose constituents are
a confined dressed-quark and confined dressed-scalar-diquark, and fixed its
parameters in a calculation of nucleon electromagnetic properties.  Herein we
employ that model in a study of a range of nucleon form factors that can be
identified with those used extensively in phenomenological $NN$ potentials
and meson-exchange models.  These calculations require knowledge of the
relevant meson Bethe-Salpeter amplitudes and $3$-point vertex functions.
However, they have been determined in the application of Dyson-Schwinger
equation models to non-nucleonic processes.  It is an important result that
this simple model provides a uniformly good description of nucleon properties
and, where there are discrepancies with experimental data, a cause and a
means for its amelioration are readily identified.  Our study demonstrates
that it is realistic to hope for useful constraints on meson-exchange models
from well-constrained models of hadron structure.

Our calculations suggest that the nucleon form factors are ``soft'' and there
is no sign that this is a model-dependent result.  The couplings generally
agree well with those fitted in meson-exchange models.  The only significant
discrepancy is that we find $0.4 \lsim f_{\omega NN}/g_{\omega NN}\lsim 1.0$,
whereas the conventional model assumption is $f_{\omega NN}=0$.  Comparison
with our calculation of the nucleon's isoscalar electromagnetic form factor,
however, suggests that $f_{\omega NN}$ is the one coupling particularly
sensitive to neglecting the axial-vector diquark.  Hence a conclusive
determination of $f_{\omega NN}$ must await its incorporation.

A primary requirement for improving our model is the inclusion of the
axial-vector diquark correlation.  In our study of $g_A$ we saw that it can
contribute up to$\,\sim 25$\%, and Fadde'ev equation studies
show\cite{ishii,raAB,regFB} that it provides a necessary $\sim 33$\%
reduction of the quark$+$scalar-diquark nucleon mass.  Also important in our
analysis of $g_A$ was an elucidation of the role played by transverse parts
of the dressed-quark-axial-vector vertex that are regular at $Q^2=0$.  A
simple model that allowed for the constrained leakage of the $a_1$-meson pole
contribution into the spacelike region showed that terms unrestricted by the
axial-vector Ward-Takahashi identity provide $\sim 20$\% of the magnitude of
$g_A$.  This sensitivity of the result to such elements makes important a
numerical solution of the axial-vector vertex equation, a calculation for
which the study of the vector vertex\cite{marispion} serves as an exemplar.

Our analysis of the nucleon $\sigma$ term is particularly interesting because
it illustrates the only method that allows an unambiguous off-shell
extrapolation in the estimation of meson-nucleon form factors.  An essential
element in the impulse approximation calculation of the scalar form factor:
$\sigma(q^2)$, is the dressed-quark-scalar vertex and we used a separable
model to obtain it as the solution of the inhomogeneous scalar vertex
equation.  This solution exhibits a simple pole at $Q^2= -m_\sigma^2$ and
hence so does $\sigma(q^2)$.  The residue of that pole gives the
$\sigma$-meson nucleon coupling.  However, the inhomogeneous vertex equation
admits a solution for arbitrary $Q^2$, which describes the $Q^2$-dependent
dressed-quark-scalar coupling and hence allows a direct and consistent
determination of $\sigma=\sigma(q^2=0)$.  That $Q^2$-dependent coupling
exemplifies the necessary elements in studies of those meson-nucleon form
factors that truly represent correlated quark exchange.  Our calculation of
$\sigma(q^2)$ and the model $b)$ calculation of $g_A$ are analogues of
Ref.~\cite{marispion}, which elucidates similar aspects of the
electromagnetic pion form factor, making explicit the $\rho$-meson
contribution and its leakage away from $Q^2=-m_\rho^2$.

As noted above, to proceed it is important to include axial-vector diquark
correlations.  Without them we can't describe the $\Delta$ resonance, and the
$N \to \Delta$ transition is an important probe of hadron structure and
models; e.g., resonant quadrupole strength in this transition can be
interpreted as a signal of nucleon deformation\cite{costas}.  The existence
of strong final state interactions muddies this interpretation and means that
nucleon structure models such as ours can't be compared directly with data.
However, they can be used as a foundation in the application of detailed
reaction models\cite{harryII} and thereby provide a connection between the
nucleon's quark-gluon content, its ``shape'' and data.\\[0.2\baselineskip]

{\bf Acknowledgments}~We acknowledge interactions with R.~Alkofer,
T.-S.H.~Lee, H.B.~O'Connell and P.C.~Tandy.  This work was supported by the
US Department of Energy, Nuclear Physics Division, under contract
no. W-31-109-ENG-38, and benefited from the resources of the National Energy
Research Scientific Computing Center.  S.M.S. is grateful for financial
support from the A.v.~Humboldt foundation.

\appendix
\section*{Collected Formulae}
\subsection{Pion-Nucleon}
For the $\pi NN$ coupling, Fig.~\ref{diagrams} represents:
\begin{eqnarray}
\label{piNN1} 
\lefteqn{\Lambda^{1j}_\pi(q,P)  =  3 \int\!\!\sfrac{d^4 \ell}{(2\pi)^4}\,
}\\ & & \nonumber \psi(K,p_3+q) \Delta(K)
\psi(K,p_3)\,\Lambda_\pi^{qj}(p_3+q,p_3) \,,
\end{eqnarray}
with 
$K= \case{2}{3} P + \ell$, $p_3= \case{1}{3} P -\ell$, $p_2= K/2 - k$,
$\Lambda_\pi^{qj}(k_1,k_2) = S(k_1)\,\Gamma_\pi^j(k_r;k_T)\,S(k_2)$, $k_r =
\case{1}{2}(k_1-k_2)$, $k_T=k_1+k_2$, 
\begin{equation}
\Lambda^{2j}_\pi(q,P)  =  0\,;
\end{equation}
i.e., there isn't a direct pion--scalar-diquark coupling because of parity
conservation, 
\begin{eqnarray}
\lefteqn{\Lambda^{3j}_\pi(q,P) = -6\int\!\!\sfrac{d^4 k}{(2\pi)^4}\sfrac{d^4
\ell}{(2\pi)^4}\,\Omega(p_1+q,p_3,p_2)\,   }\\ && \nonumber \times
\Omega(p_1,p_2,p_3)\, S(p_2) \, \Lambda^{qj}_\pi(p_1,p_1+q)\,S(p_3)\,,\\
\lefteqn{\Lambda^{4j}_\pi(q,P) = 6\int\sfrac{d^4 k}{(2\pi)^4}\sfrac{d^4
\ell}{(2\pi)^4}\,\Omega(p_1,p_3,p_2+q)\,  }\\ && \nonumber \times
\Omega(p_1,p_2,p_3) \, \Lambda^{qj}_\pi(p_2+q,p_2) \,S(p_1) \, S(p_3)\,,\\
\label{piNN5} \lefteqn{\Lambda^{5j}_\pi(q,P) = 6\int\sfrac{d^4
k}{(2\pi)^4}\sfrac{d^4 \ell}{(2\pi)^4}\, \Omega(p_1,p_3+q,p_2)\,  }\\ &&
\nonumber \times \Omega(p_1,p_2,p_3)\,
S(p_2)\,S(p_1)\,\Lambda^{qj}_\pi(p_3+q,p_3) \,,
\end{eqnarray}
with $6 = \varepsilon_{c_1 c_2 c_3} \varepsilon_{c_1 c_2 c_3}$ and
\begin{equation}
\Omega(p_1,p_2,p_3) = \psi(p_1+p_2,p_3) \,\Delta(p_1+p_2) \, \Gamma(p_1,p_2)\,.
\end{equation}

\subsection{Vector-meson--Nucleon}
For the vector meson coupling, Fig.~\ref{diagrams} represents:
\begin{eqnarray}
\label{V1}
\lefteqn{\Lambda^{1j}_\mu(q,P)  = 3 \int\!\!\sfrac{d^4 \ell}{(2\pi)^4}\, }\\ &
& \nonumber \psi(K,p_3+q) \Delta(K) \psi(K,p_3)\,\Lambda_\mu^{qj}(p_3+q,p_3)
\,, \\
\lefteqn{\Lambda^{2j}_\mu(q,P) = 6 \int\!\!\sfrac{d^4 k}{(2\pi)^4}\sfrac{d^4
\ell}{(2\pi)^4}\, \Omega(p_1+q,p_2,p_3)\, }\\ && \nonumber \times
\Omega(p_1,p_2,p_3) \, {\rm tr}_{DF}\left[ \Lambda_\mu^{qj}(p_1+q,p_1)
S(p_2)\right]\, S(p_3)\,,\\
\lefteqn{\Lambda^{3j}_\mu(q,P) = (-1)^I \,6\int\!\!\sfrac{d^4
k}{(2\pi)^4}\sfrac{d^4 \ell}{(2\pi)^4}\,\Omega(p_1+q,p_3,p_2)\,   }\\ &&
\nonumber \times \Omega(p_1,p_2,p_3)\, S(p_2) \,
\Lambda^{qj}_\mu(p_1,p_1+q)\,S(p_3)\,,\\
\lefteqn{\Lambda^{4j}_\mu(q,P) = 6\int\sfrac{d^4 k}{(2\pi)^4}\sfrac{d^4
\ell}{(2\pi)^4}\,\Omega(p_1,p_3,p_2+q)\,  }\\ && \nonumber \times
\Omega(p_1,p_2,p_3) \, \Lambda^{qj}_\mu(p_2+q,p_2) \,S(p_1) \, S(p_3)\,,\\
\label{V5}
\lefteqn{\Lambda^{5j}_\mu(q,P) = 6\int\sfrac{d^4 k}{(2\pi)^4}\sfrac{d^4
\ell}{(2\pi)^4}\, \Omega(p_1,p_3+q,p_2)\,  }\\ && \nonumber \times
\Omega(p_1,p_2,p_3)\, S(p_2)\,S(p_1)\,\Lambda^{qj}_\mu(p_3+q,p_3) \,,
\end{eqnarray}
with $\Lambda_\mu^{qj}(k_1,k_2) = S(k_1)\,\Gamma_\mu^j(k_r;k_T)\,S(k_2)$ and
$\Gamma_\mu^j(k_r;k_T)$ the Bethe-Salpeter-like amplitude of Eq.~(\ref{BSAV})
whose normalisation is determined by
\begin{eqnarray}
\label{Vnorm}
\lefteqn{2\,{\cal N}_V^2\, \delta^{\alpha\beta} Q_\mu = \case{1}{3}
\mbox{tr}_{\rm CDF} \int \frac{d^4 k}{(2\pi)^4} } \\
&&\left[\Gamma_\nu^\alpha(k;-Q)\, \frac{\partial S(k_+)}{\partial
Q_\mu}\,\Gamma_\nu^\beta(k;Q) \,S(k_-) \right. \nonumber
\\&& \left. + \Gamma_\nu^\alpha(k;-Q)\, S(k_+)\,\Gamma_\nu^\beta(k;Q)\,
\frac{\partial S(k_-)}{\partial Q_\mu}\right]_{Q^2=-M_V^2} \nonumber \,,
\end{eqnarray}
where $k_{\pm} = k \pm Q/2$.  The flavour trace ensures that
$\Lambda^{2j}_\mu$ contributes only to the isoscalar coupling.  This merely
reflects the fact that in an isospin symmetric theory there can't be a
three-point iso-vector-scalar-scalar vertex.  Further, $\Lambda^{3j}_\mu$
contributes with opposite signs to the $\omega^{(I=0)}$ and $\rho^{(I=1)}$
couplings.  The complete vertex is
\begin{equation}
\label{fullvector}
\Lambda^j_{\mu}(q,P) = \Lambda^{1j}_\mu(q,P) + 
2 \sum_{n=2}^5\,\Lambda^{nj}_\mu(q,P)\,.
\end{equation}

\subsection{Axial-vector--Nucleon}
Model $b)$ for the dressed-quark-axial-vector vertex involves the
Bethe-Salpeter amplitude for the $a_1$ meson, which in the separable model of
Ref.~\cite{bsesep} has the form (terms quadratic in $k$ are suppressed in
this model)
\begin{equation}
\label{BSAa1}
\Gamma_\mu^{a_1}(\bar k,\hat Q) = i\vec{\tau} \left[
\gamma_5\gamma_\mu^T e_1^{a_1} \bar G(x)
+ e_2^{a_1}
\epsilon_{\lambda\mu\nu\sigma}\gamma_\lambda\bar k_\nu\hat Q_\sigma \bar
F(x)
\right]
\end{equation}
where $\bar k = k/\lambda$, $\lambda =0.566\,$GeV is the model's mass-scale,
$\hat Q_\mu = Q_\mu/|Q^2|^{1/2}$, $ \gamma_\mu^T= (\gamma_\mu - Q_\mu
\gamma\cdot Q /Q^2)$ and
\begin{equation}
\label{FGa1}
\bar F(x) = \frac{1}{\bar a}\left[ \bar A(x) - 1 \right]\,,\;
\bar G(x) = \frac{1}{\bar b}\left[ \bar B(x) - \bar m \right]\,,
\end{equation}
with calculated constants
\begin{equation}
\bar a= (0.359)^2\,,\;\bar b=(0.296)^2\,.
\end{equation}
Herein\cite{a1b1} $\bar A(x)$, $\bar B(x)$ are modified dressed-quark
propagator functions obtained in the usual way from
\begin{eqnarray}
\label{ssa1}
\bar\sigma_S^{a_1}(x)&  = &
2\,\hat m \,{\cal F}(2 (x+ \hat m^2))\\
&& \nonumber
+ {\cal F}(b_1 x) \,{\cal F}(b_3 x) \,{\cal F}((\epsilon_S x)^2)\,
\left[b_0 + b_2^{a_1} {\cal F}(\epsilon x)\right]\,,\\
\label{sva1}
\bar\sigma_V^{a_1}(x) & = & 
\frac{ 2 \,(x+\hat m^2) - {\rm e}^{-\epsilon_V^2 (x+\hat m^2)^2}
        + {\rm e}^{-2 (x+\hat m^2)}}{2\, (x+\hat m^2)^2}
\end{eqnarray}
where 
\begin{equation}
\label{sepa1}
\hat m =0.0081\,,\; b_2^{a_1}= 0.863\,,\;
\epsilon_S = 0.482\,,\;\epsilon_V=0.1\,,
\end{equation}
and the other parameters are given in Eq.~(\ref{tableA}).

The separable model yields $m_{a_1}$ in Eq.~(\ref{a1res}) and the eigenvector
\begin{equation}
e_1^{a_1} = 0.145\,,\;
e_2^{a_1} = 1.69\,,
\end{equation}
which is canonically normalised using Eq.~(\ref{Vnorm}) evaluated with
Eqs.~(\ref{BSAa1}-\ref{sepa1}).  This Bethe-Salpeter amplitude yields
$f_{a_1}$ in Eq.~(\ref{a1res}).
\begin{equation}
\label{a1res} 
\begin{array}{l|cc}
           & m_{a_1}\,({\rm GeV}) &  f_{a_1}\,({\rm GeV}) \\\hline
{\rm Calc.}& 1.34  & 0.221  \\
{\rm Obs.}
           & 1.23 \pm 0.040~\protect\cite{pdg}
           & 0.203\pm 0.018~\protect\cite{isgur} 
\end{array}
\end{equation}
NB: This model predicts that the $\gamma_5\gamma_\mu$ term is sub-dominant in
the $a_1$ meson.  The dominant $e_2^{a_1}$-term characterises constituents
with relative orbital motion.

For the axial-vector coupling, Fig.~\ref{diagrams} represents:
\begin{eqnarray}
\label{AV1}
\lefteqn{\Lambda^{1j}_{5\mu}(q,P)  = 3 \int\!\!\sfrac{d^4 \ell}{(2\pi)^4}\, }\\
& & \nonumber \psi(K,p_3+q) \Delta(K)
\psi(K,p_3)\,\Lambda_{5\mu}^{qj}(p_3+q,p_3) \,, \\
\lefteqn{\Lambda^{2j}_{5\mu}(q,P) = 0} \,,\\
\lefteqn{\Lambda^{3j}_{5\mu}(q,P) = 6 \int\!\!\sfrac{d^4 k}{(2\pi)^4}\sfrac{d^4
\ell}{(2\pi)^4}\,\Omega(p_1+q,p_3,p_2)\,   }\\ && \nonumber \times
\Omega(p_1,p_2,p_3)\, S(p_2) \, \Lambda^{qj}_{5\mu}(p_1,p_1+q)\,S(p_3)\,,\\
\lefteqn{\Lambda^{4j}_{5\mu}(q,P) = 6\int\sfrac{d^4 k}{(2\pi)^4}\sfrac{d^4
\ell}{(2\pi)^4}\,\Omega(p_1,p_3,p_2+q)\,  }\\ && \nonumber \times
\Omega(p_1,p_2,p_3) \, \Lambda^{qj}_{5\mu}(p_2+q,p_2) \,S(p_1) \, S(p_3)\,,\\
\label{AV5}
\lefteqn{\Lambda^{5j}_{5\mu}(q,P) = 6\int\sfrac{d^4 k}{(2\pi)^4}\sfrac{d^4
\ell}{(2\pi)^4}\, \Omega(p_1,p_3+q,p_2)\,  }\\ && \nonumber \times
\Omega(p_1,p_2,p_3)\, S(p_2)\,S(p_1)\,\Lambda^{qj}_{5\mu}(p_3+q,p_3) \,,
\end{eqnarray}
$\Lambda_{5\mu}^{qj}(k_1,k_2) = S(k_1)\,\Gamma_{5\mu}^j(k_r;k_T)\,S(k_2)$
with $\Gamma_{5\mu}^j(k_r;k_T)= \Gamma_{5\mu}^{Rj}(k_r;k_T)$, the regular
part of the dressed-axial-vector-quark vertex defined by: model $a)$,
Eqs.~(\ref{AVAnsatz},\ref{AVVa}); or model $b)$,
Eqs.~(\ref{AVAnsatz},\ref{AVVb}).  $\Lambda^{2j}_\mu$ vanishes for the same
reason that $\Lambda^{2j}_\pi$ does and the complete vertex is
\begin{equation}
\label{fullaxial}
\Lambda^j_{5\mu}(q,P) = \Lambda^{1j}_{5\mu}(q,P) + 2
\sum_{n=2}^5\,\Lambda^{nj}_{5\mu}(q,P)\,.
\end{equation}

\subsection{Sigma term}
The contributions to the scalar-nucleon vertex are:
\begin{eqnarray}
\label{S1}
\lefteqn{\Lambda^{1}_{\mbox{\large\boldmath $1$}}(q,P) = 3 \int\!\!\sfrac{d^4
\ell}{(2\pi)^4}\, }\\ & & \nonumber \psi(K,p_3+q) \Delta(K)
\psi(K,p_3)\,\Lambda_{m}^{q}(p_3+q,p_3) \,, \\
\lefteqn{\Lambda^{2}_{\mbox{\large\boldmath $1$}}(q,P) = 12
\int\!\!\sfrac{d^4 k}{(2\pi)^4}\sfrac{d^4 \ell}{(2\pi)^4}\,
\Omega(p_1+q,p_2,p_3)\, }\\ && \nonumber \times \Omega(p_1,p_2,p_3) \, {\rm
tr}_{D}\left[ \Lambda_{m}^{q}(p_1+q,p_1) S(p_2)\right]\, S(p_3)\,,\\
\lefteqn{\Lambda^{3}_{\mbox{\large\boldmath $1$}}(q,P) = 6 \int\!\!\sfrac{d^4
k}{(2\pi)^4}\sfrac{d^4 \ell}{(2\pi)^4}\,\Omega(p_1+q,p_3,p_2)\, }\\ &&
\nonumber \times \Omega(p_1,p_2,p_3)\, S(p_2) \,
\Lambda^{q}_{m}(p_1,p_1+q)\,S(p_3)\,,\\
\lefteqn{\Lambda^{4}_{\mbox{\large\boldmath $1$}}(q,P) = 6\int\sfrac{d^4
k}{(2\pi)^4}\sfrac{d^4 \ell}{(2\pi)^4}\,\Omega(p_1,p_3,p_2+q)\, }\\ &&
\nonumber \times \Omega(p_1,p_2,p_3) \, \Lambda^{q}_{m}(p_2+q,p_2) \,S(p_1)
\, S(p_3)\,,\\
\label{S5}
\lefteqn{\Lambda^{5}_{\mbox{\large\boldmath $1$}}(q,P) = 6\int\sfrac{d^4
k}{(2\pi)^4}\sfrac{d^4 \ell}{(2\pi)^4}\, \Omega(p_1,p_3+q,p_2)\, }\\ &&
\nonumber \times \Omega(p_1,p_2,p_3)\,
S(p_2)\,S(p_1)\,\Lambda^{q}_{m}(p_3+q,p_3) \,,
\end{eqnarray}
with $\Lambda_{m}^{q}(k_1,k_2) = S(k_1)\,\Gamma_{m}(k_r;k_T)\,S(k_2)$, and
\begin{equation}
\label{fullscalar}
\Lambda_{\mbox{\large\boldmath $1$}}(q,P) =
\Lambda^{1}_{\mbox{\large\boldmath $1$}}(q,P) + 2
\sum_{n=2}^5\,\Lambda^{n}_{\mbox{\large\boldmath $1$}}(q,P)\,.
\end{equation}



\end{document}